\documentstyle [12pt] {article}

\catcode`@=12
\begin{document}
\thispagestyle{empty}
\begin{flushright} UCRHEP-T263\\September 1999\
\end{flushright}
\vskip 0.5in
\begin{center}
{\Large \bf Fitting Precision Electroweak Data\\
with Exotic Heavy Quarks\\}
\vskip 1.0in
{\bf Darwin Chang, We-Fu Chang\\}
\vskip 0.1in
{\sl NCTS and Department of Physics, National Tsing-Hua University,\\}
{\sl Hsinchu 30043, Taiwan, ROC\\}
\vskip 0.5in
{\bf Ernest Ma\\}
\vskip 0.1in
{\sl Physics Department, Univeristy of California,\\}
{\sl Riverside, CA 92521, USA\\}
\end{center}
\vskip 1.0in
\begin{abstract}\
The 1999 precision electroweak data from LEP and SLC persist in showing 
some slight discrepancies from the assumed standard model, mostly 
regarding $b$ and $c$ quarks.  We show how their mixing with exotic heavy 
quarks could result in a more consistent fit of all the data, including 
two unconventional interpretations of the top quark.
\end{abstract}
\newpage
\baselineskip 24pt
Precision measurements of electroweak parameters at the $Z$ resonance have 
been available for many years\cite{1,2}.  Their updated values in 1999 as 
reported at Tampere\cite{3} and at Stanford\cite{4} are consistent with the 
expectations of the minimal standard model, including all radiative 
corrections to one-loop order.  However, certain slight discrepancies 
persist, mostly regarding $b$ and $c$ quarks.  In this note, we show 
how their mixing with exotic heavy quarks could result in a more consistent 
fit of all the data, including two unconventional interpretations of the 
top quark.

The most telling sign that there may be something beyond the minimal 
standard model in precision electroweak measurements is the 
observation\cite{4} that the two most precise measurements of 
$\sin^2 \theta_{eff}$ are 3.0 standard deviations apart.  One is the 
left-right asymmetry $A_{LR}$ (which directly measures $A_e$) from SLC 
at SLAC that gives\cite{4}
\begin{equation}
\sin^2 \theta_{eff} (A_{LR}) = 0.23101 \pm 0.00028,
\end{equation}
and the other is the forward-backward asymmetry $A^{0,b}_{FB}$ of $b$ quarks 
from LEP at CERN which gives\cite{3}
\begin{equation}
\sin^2 \theta_{eff} (A^{0,b}_{FB}) = 0.23236 \pm 0.00036.
\end{equation}
We note that Eq.~(1) is consistent with the forward-backward asymmetry of 
leptons measured at LEP which gives\cite{3}
\begin{equation}
\sin^2 \theta_{eff} (A^{0,l}_{FB}) = 0.23107 \pm 0.00053,
\end{equation}
whereas Eq.~(2) is consistent with the $A_b$ measurement at SLC, 
i.e. $A_b = 0.905 \pm 0.026$ versus the extracted value\cite{4} of $A_b = 
0.881 \pm 0.020$ from the value of $A^{0,b}_{FB}$ shown.  This points 
to the possibility that there is new physics in the decay $Z \to b \bar b$. 

Specifically, consider the effective left-handed and right-handed couplings 
of the $b$ quark to the $Z$ boson in the standard model:
\begin{eqnarray}
g_{bL}^{SM} &=& \left( 1 + {\epsilon_1 \over 2} \right) \left( - {1 \over 2} 
(1 + \epsilon_b) + {1 \over 3} \sin^2 \theta_{eff} \right), \\ 
g_{bR}^{SM} &=& \left( 1 + {\epsilon_1 \over 2} \right) {1 \over 3} \sin^2 
\theta_{eff},
\end{eqnarray}
where the radiative corrections\cite{2} $\epsilon_1$ and $\epsilon_b$ are 
functions of $m_t$ and $m_H$.  Note the important fact\cite{5} that 
$\epsilon_b$ (which has a strong quadratic dependence on $m_t$) contributes 
only to $g_{bL}^{SM}$.  On the other hand, the measured quantity 
$R_b \equiv \Gamma (Z \to b \bar b) / \Gamma (Z \to {\rm hadrons})$ is 
proportional to $g_{bL}^2 + g_{bR}^2$, whereas $A_{FB}^{0,b}$ and $A_b$ 
are proportional to $(g_{bL}^2 - g_{bR}^2)/(g_{bL}^2 + g_{bR}^2)$.  From the 
1999 data reported at Tampere\cite{3} and at Stanford\cite{4},
\begin{equation}
R_b = 0.21642 \pm 0.00073, ~~~ A^{0,b}_{FB} = 0.0984 \pm 0.0020, ~~~ 
A_b = 0.905 \pm 0.026,
\end{equation}
the couplings $g_{bL}$ and $g_{bR}$
can be extracted\cite{6}:
\begin{equation}
g_{bL} = -0.4163 \pm 0.0020, ~~~ g_{bR} = 0.0996 \pm 0.0076.
\end{equation}
Using $m_t = 174$ GeV, $m_H = 100$ GeV, and $\alpha(m_Z)^{-1}=128.9$,
the standard model yields\cite{7}
\begin{equation}
g_{bL}^{SM} = -0.4208, ~~~ g_{bR}^{SM} = 0.0774.
\end{equation}
Note that $g_{bL}^2 + g_{bR}^2$ is almost exactly equal to $(g_{bL}^{SM})^2 
+ (g_{bR}^{SM})^2$, but $g_{bL}$ and $g_{bR}$ are each over two standard 
deviations away from $g_{bL}^{SM}$ and $g_{bR}^{SM}$ respectively.

As we already pointed out last year\cite{5}, since $\epsilon_b$ depends only 
on the left-handed partner of the $b$ quark, this may be an indication that 
$m_t$ is actually much greater than 174 GeV and the observed ``top'' events 
are due to an exotic quark $Q_4$ of charge $-4/3$.  In this scenario, the 
singlet $b_R$ mixes with the exotic quark $Q_1$ in the doublet $(Q_1, Q_4)_R$ 
so that
\begin{eqnarray}
g_{bR} &=& \left( 1 + {\epsilon_1 \over 2} \right) \left[ {1 \over 3} \sin^2 
\theta_{eff} \cos^2 \theta_b + \left( {1 \over 2} + {1 \over 3} \sin^2 
\theta_{eff} \right) \sin^2 \theta_b \right] \nonumber \\ &=& \left( 1 + 
{\epsilon_1 \over 2} \right) \left( {1 \over 3} \sin^2 \theta_{eff} + 
{1 \over 2} \sin^2 \theta_b \right).
\end{eqnarray}
Since $\sin^2 \theta_{eff}/3$ is small to begin with, a reasonably small 
$\sin^2 \theta_b$ is sufficient to make $g_{bR}$ fit the data.  [If radiative 
corrections to $g_{bR}$ from new physics were
invoked, an unreasonably large effect of 
about 30\% would be needed.]  In the following we will update our analysis 
using the 1999 data.  We will also address the new possibility that slight 
discrepancies in $Z \to c \bar c$ may be due to yet another exotic 
quark\cite{8} and offer a second alternative interpretation of the ``top'' 
events.

Using the 1999 $Z \to l^- l^+$ data assuming lepton
universality\cite{3,4}, i.e.
\begin{equation}
\Gamma_l = 83.96 \pm 0.09 ~{\rm MeV}, ~~~ A^{0,l}_{FB} = 0.01701 \pm 0.00095,
\end{equation}
together with\cite{3}
\begin{equation}
m_W = 80.394 \pm 0.042 ~{\rm GeV}, ~~~ m_Z = 91.1871 \pm 0.0021 ~{\rm GeV},
\end{equation}
we find
\begin{equation}
\epsilon_1 = (4.7 \pm 1.1) \times 10^{-3}, ~~~ \epsilon_2 = (-7.2 \pm 2.4) 
\times 10^{-3}, ~~~ \epsilon_3 = (3.6 \pm 1.7) \times 10^{-3},
\end{equation}
which agree very well with previous values\cite{2,5} and also with the 
standard model, i.e.\cite{6}
\begin{equation}
\epsilon_1^{SM} = 5.4 \times 10^{-3}, ~~~ \epsilon_2^{SM} = -7.6 \times 
10^{-3}, ~~~ \epsilon_3^{SM} = 5.2 \times 10^{-3}.
\end{equation}
Using Eqs.~(3), (4) and (7), we then obtain
\begin{equation}
\epsilon_b = (-15.3 \pm 4.0) \times 10^{-3}.
\end{equation}
This implies that
\begin{equation}
m_t = 271 \begin{array}{l} +33 \\ -38 \end{array} {\rm GeV},
\end{equation}
where we have approximated $\epsilon_b$ by its leading contribution, $-G_F 
m_t^2/4 \pi^2 \sqrt 2$.  To explain $g_{bR}$ of Eq.~(7) and thus also Eq.~(2), 
we use Eq.~(9) and find
\begin{equation}
\sin^2 \theta_b = 0.045 \pm 0.015.
\end{equation}
In the standard model, $\epsilon_1$ and $\epsilon_b$ are fixed by 
$m_t = 174$ GeV and $\theta_b$ is absent, so the experimental discrepancy 
from $Z \to b \bar b$ data is forced into a value of $\sin^2 \theta_{eff}$ 
given by Eq.~(2) which is 3.0 standard deviations away from the true 
value given by Eqs.~(1) and (3).

Our interpretation of the data so far is that $b_R$ is not purely $I_3 = 0$ 
as in the standard model, but has a small $I_3 = 1/2$ component from mixing 
with the exotic $(Q_1, Q_4)_R$ doublet.  We also take the viewpoint that 
$b_L$ is as given by the standard model and the measured $g_{bL}$ is a 
direct indication of the mass of its partner, defined as the $t$ quark. 
This results in Eq.~(15).  At this point, we need to revise our assessment 
of the agreement of Eq.~(12) with Eq.~(13), namely that in the presence of 
new physics, $\epsilon_{1,2,3}$ receive additional contributions, hence a 
change in the value of $m_t$ may be suitably compensated.  Details have 
already been discussed in our previous paper\cite{5}.

Consider now the 1999 $Z \to c \bar c$ data:
\begin{equation}
R_c = 0.1674 \pm 0.0038, ~~~ A^{0,c}_{FB} = 0.0691 \pm 0.0037, ~~~ 
A_c = 0.630 \pm 0.026,
\end{equation}
from which the couplings $g_{cL}$ and $g_{cR}$ can be extracted\cite{6}:
\begin{equation}
g_{cL} = 0.341 \pm 0.005, ~~~ g_{cR} = -0.164 \pm 0.005,
\end{equation}
whereas the standard model yields\cite{6}
\begin{equation}
g_{cL}^{SM} = 0.347, ~~~ g_{cR}^{SM} = -0.155.
\end{equation}
Although the deviations here are small, there is a hint that $g_{cR}$ may 
be too big in magnitude and $g_{cL}$ too small.  To explain both, we take 
the analog of Eq.~(9) and let $c$ mix with a heavy quark $Q_2$, where 
$Q_{2L}$ is a singlet but $(Q_5, Q_2)_R$ is an exotic doublet, so that
\begin{eqnarray}
g_{cR} &=& \left( 1 + {\epsilon_1 \over 2} \right) \left( -{2 \over 3} \sin^2 
\theta_{eff} - {1 \over 2} \sin^2 \theta_{cR} \right), \\ g_{cL} &=& \left( 
1 + {\epsilon_1 \over 2} \right) \left( {1 \over 2} - {2 \over 3} \sin^2 
\theta_{eff} - {1 \over 2} \sin^2 \theta_{cL} \right).
\end{eqnarray}
Using Eqs.~(3), (12) and (18), we then obtain
\begin{equation}
\sin^2 \theta_{cR} = 0.02 \pm 0.01, ~~~ \sin^2 \theta_{cL} = 0.01 \pm 0.01.
\end{equation}
This opens up the possibility that $Q_2$ may also mix with $t$ (and not just 
with $c$) so that the Tevatron ``top'' events are due to $Q_2$ rather than 
$t$ which is heavier.  This second interpretation is of course much more 
speculative because it is not directly related to the data.  Note that 
the $\epsilon_{1,2,3}$ contributions of $Q_2$ and $Q_5$ may be handled in 
the same way as those of $Q_1$ and $Q_4$, as discussed by us in Ref.~[5].

In conclusion, we have shown in this short note that the 1999 precision 
electroweak data at LEP and SLC still support the possibility\cite{5} that 
$b_R$ mixes with $Q_{1R}$ of the exotic heavy quark doublet $(Q_1, Q_4)_R$. 
Hence the ``top'' events may be due to $Q_4$ which has charge $-4/3$, whereas 
the true $t$ quark is heavier, as evidenced by the value of $\epsilon_b$ 
extracted from $g_{bL}$.  Experimentally, $t \to b W^+$ and $\bar Q_4 \to 
\bar b W^+$ are not distinguishable at the Tevatron at present because the 
$b$ or $\bar b$ jet charge is not easily measured, but that will become 
possible in the near future.  We also propose here a second, more 
speculative idea that the ``top'' events may be due to a heavy quark $Q_2$ 
of charge 2/3, where $Q_{2L}$ is a singlet but $(Q_5, Q_2)_R$ is an 
exotic doublet.  In both scenarios, the lifetime of the ``top'' is 
enhanced by the inverse square of a reduced coupling and the single 
production of ``top'' at the Tevatron is suppressed.
\vskip 0.5in
\begin{center} {ACKNOWLEDGEMENT}
\end{center}

We thank J. H. Field and P. B. Renton for valuable correspondence. 
This work was supported in part by the U.~S.~Department of Energy under Grant 
No.~DE-FG03-94ER40837 and a grant from the National Science Council of the 
Republic of China.

\newpage
\bibliographystyle{unsrt}

\end{document}